\documentclass[aps,prd,showpacs,showkeys,
preprintnumbers,unsortedaddress,superscriptaddress,
nofootinbib,notitlepage]{revtex4-1}

\usepackage[T1]{fontenc}
\usepackage[utf8]{inputenc}
\usepackage{amsmath}
\usepackage{amssymb}
\usepackage{hyperref}
\usepackage{graphicx} 

\begin{document}

\title{Towards realistic models of quark masses with geometrical CP violation}
\author{Ivo de Medeiros Varzielas}
\email{ivo.de@unibas.ch}
\affiliation{\small Facult\"at f\"ur Physik, Technische 
      Universit\"at Dortmund\\ D-44221 Dortmund, Germany}
\affiliation{\small Department of Physics, University of Basel,\\ Klingelbergstr. 82, CH-4056 Basel, Switzerland}
\author{Daniel Pidt}
\email{daniel.pidt@tu-dortmund.de}
\affiliation{\small Facult\"at f\"ur Physik, Technische 
      Universit\"at Dortmund\\ D-44221 Dortmund, Germany}


 

\preprint{DO-TH-13/18}

\begin{abstract}
We present a model for quark masses and mixing, featuring geometrical CP violation through a $\Delta(27)$ triplet. By employing a single $U(1)_F$
or $Z_N$ symmetry in addition to $\Delta(27)$, we forbid all terms in the scalar potential that would spoil the calculable phases the triple acquires. The quark sector is realised by mimicking an existing scheme that reproduces the masses and CKM mixing, with the extra symmetry enabling the hierarchies in the Yukawa couplings through a Froggatt-Nielsen mechanism.
\end{abstract}

\maketitle


Many years ago, the concept of geometrical CP violation (GCPV) was introduced in \cite{Branco:1983tn}. This appealing particular case of spontaneous CP violation has received further attention recently \cite{deMedeirosVarzielas:2011zw, Varzielas:2012nn, Varzielas:2012pd, Bhattacharyya:2012pi, Ivanov:2013nla, Varzielas:2013zbp, Ma:2013xqa}.
In general, there has been lot of interest in the study of CP violation in the context of discrete flavour groups, with CP symmetries  now being frequently employed together with flavour groups in the context of models explaining fermion masses and mixing 
\cite{Chen:2009gf, Branco:2012vs, Ahn:2012tv, Meroni:2012ty, Mohapatra:2012tb, Feruglio:2012cw, Holthausen:2012dk, Antusch:2013wn, Ding:2013hpa, Feruglio:2013hia, Antusch:2013kna, Nishi:2013jqa, Luhn:2013lkn, Antusch:2013tta, Ahn:2013eua}.

As shown originally in \cite{Branco:1983tn}, in a multi-Higgs extension of the Standard Model (SM) based on $\Delta(27)$ commuting with a CP symmetry, the renormalisable scalar potential can lead to spontaneous CP violation through a complex vacuum expectation value (VEV) of the type:\footnote{For considerations on the stability of the calculable phase under non-renormalisable terms in the scalar potential see \cite{Varzielas:2012nn}. \cite{Holthausen:2012dk} also discusses the $\Delta(27)$ case. For generalisations of the framework to different groups and calculable phases, see \cite{Varzielas:2012pd, Ivanov:2013nla}.}
\begin{equation}
\langle H_i \rangle = v(\omega,1,1) \,,
\label{eq:VEV}
\end{equation}
with $\omega \equiv e^{i 2 \pi/3}$.
It is non-trivial to use this $\Delta(27)$ triplet VEV to produce viable patterns of fermion masses and mixing: the only parameter coming from this symmetrical VEV is the magnitude $v$ (shared by each component), and the calculable phase is very sensitive to extensions of the scalar content.
Nevertheless, after promising leading order structures were identified in \cite{deMedeirosVarzielas:2011zw}, a viable GCPV framework was finally presented explicitly for quarks in \cite{Bhattacharyya:2012pi}.
A possible approach dealing only with leptons has been more recently proposed by \cite{Ma:2013xqa}.

While minimal in terms of content, \cite{Bhattacharyya:2012pi} is not entirely satisfactory as (like the SM itself) it does not provide an explanation for the large hierarchies present in the Yukawa couplings.
Another issue is that in addition to the $\Delta(27)$ triplet scalar, at least one scalar transforming as a non-trivial singlet under $\Delta(27)$ is required to construct viable mass structures (namely, to account for the CKM mixing).
The presence of non-trivial scalar singlets affects the scalar potential: unless couplings mixing it with the triplet are assumed to be negligible, GCPV is spoiled.
While neglecting the couplings of that type increases the (accidental) symmetry of the potential and may be considered natural in some sense, the \cite{Bhattacharyya:2012pi} model did not have any symmetry beyond $\Delta(27)$ that enforced this.\footnote{This is also the case in \cite{Ma:2013xqa}, which actually has two non-trivial $\Delta(27)$ singlet scalars.} 

In order to address both of these issues of the \cite{Bhattacharyya:2012pi} model, we propose to increase the symmetry content with either a continuous $U(1)_F$ or a discrete $Z_N$ subgroup with the aim to improve upon the existing viable model.
The Yukawa hierarchies can be readily alleviated by assigning different charges to the lighter generations and implementing a Froggatt-Nielsen (FN) mechanism \cite{Froggatt:1978nt}. Furthermore, by suitably charging all non-trivial $\Delta(27)$ singlet scalars under this extra symmetry, it is possible to forbid the problematic scalar couplings.
We introduce two fields $\varphi$ and $\theta$ (SM gauge singlets), respectively charged under $\Delta(27)$ and $U(1)_F$ as $1_{00}$, $-1$ and $1_{02}$, $-2$ (see Table \ref{ta:content}).
 
\begin{table}
  \centering
  \begin{tabular}{|c||ccccc|ccc|}
    \hline
    & $Q_1$ & $Q_2$ & $Q_3$ & $u^c$ & $d^c$ & $H$ & $\varphi$ & $\theta$ \\
    \hline\hline
    $\Delta(27)$  & $1_{00}$ & $1_{00}$ & $1_{02}$ & $3_{01}$ & $3_{02}$ & $3_{01}$ & $1_{00}$ & $1_{02}$ \\
    $U(1)_F$ or $Z_N$ & $3$ & $2$ & $0$ & $0$ & $0$ & $0$ & $-1$ & $-2$ \\
    \hline
  \end{tabular}
  \caption{The symmetry and field content of the model. \label{ta:content}}
\end{table}


We implement the quark sector aiming to obtain the same $M M^\dagger$ structures for the up and down quark sectors that were shown to be viable in \cite{Bhattacharyya:2012pi}.
In the solution we propose, the $M_u$ and $M_d$ structures are not the same as in \cite{Bhattacharyya:2012pi}, but they enable the same relevant entries in $M_d M_d^\dagger$ and are therefore equally viable in terms of obtaining the CKM mixing.
Specifically, as shown in Table \ref{ta:content}, we preserve most of the $\Delta(27)$ assignments of \cite{Bhattacharyya:2012pi}: $Q_i$ (the 3 generations of quark $SU(2)$ doublets) are $1_{jk}$ singlets whereas the RH quark singlets $u^c$ and $d^c$ transform as the appropriate 3-dimensional irreducible representation that allows an invariant Yukawa coupling with $H^\dagger$ or $H$ respectively.

\begin{figure}[h!]
\centering
\includegraphics[width=5 cm]{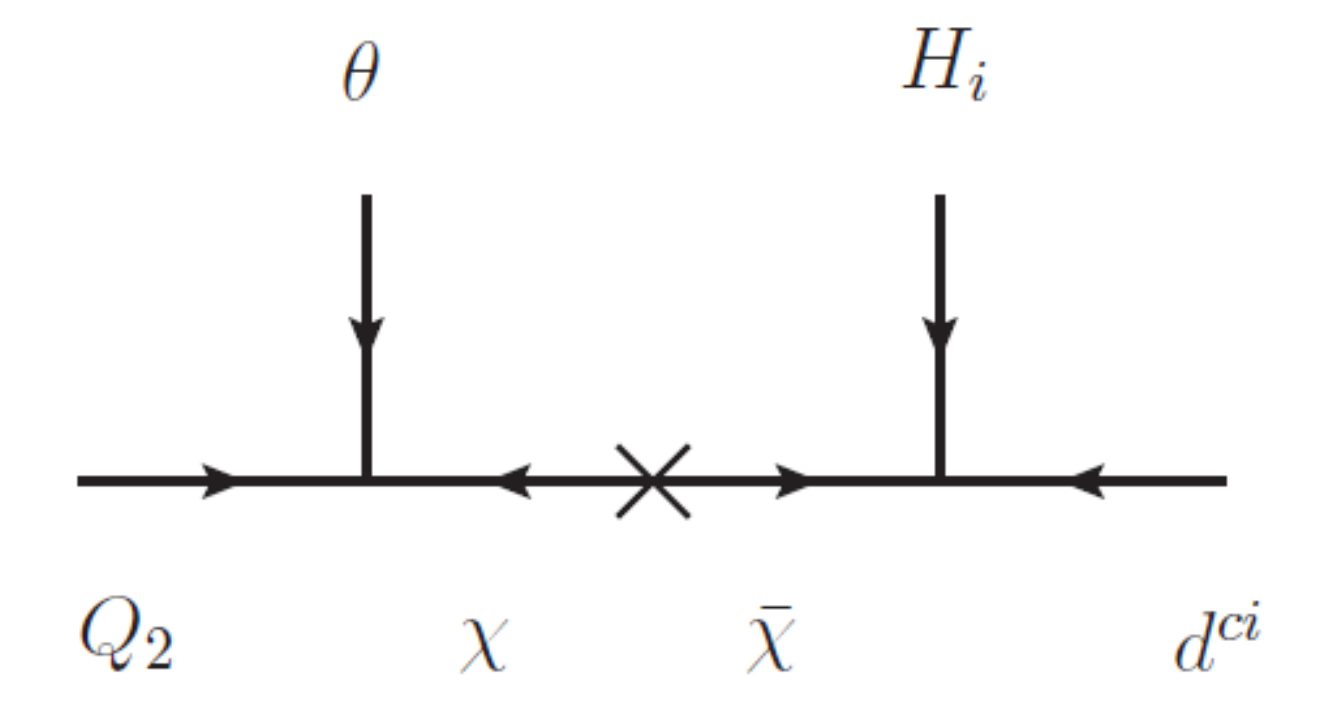}
\includegraphics[width=5 cm]{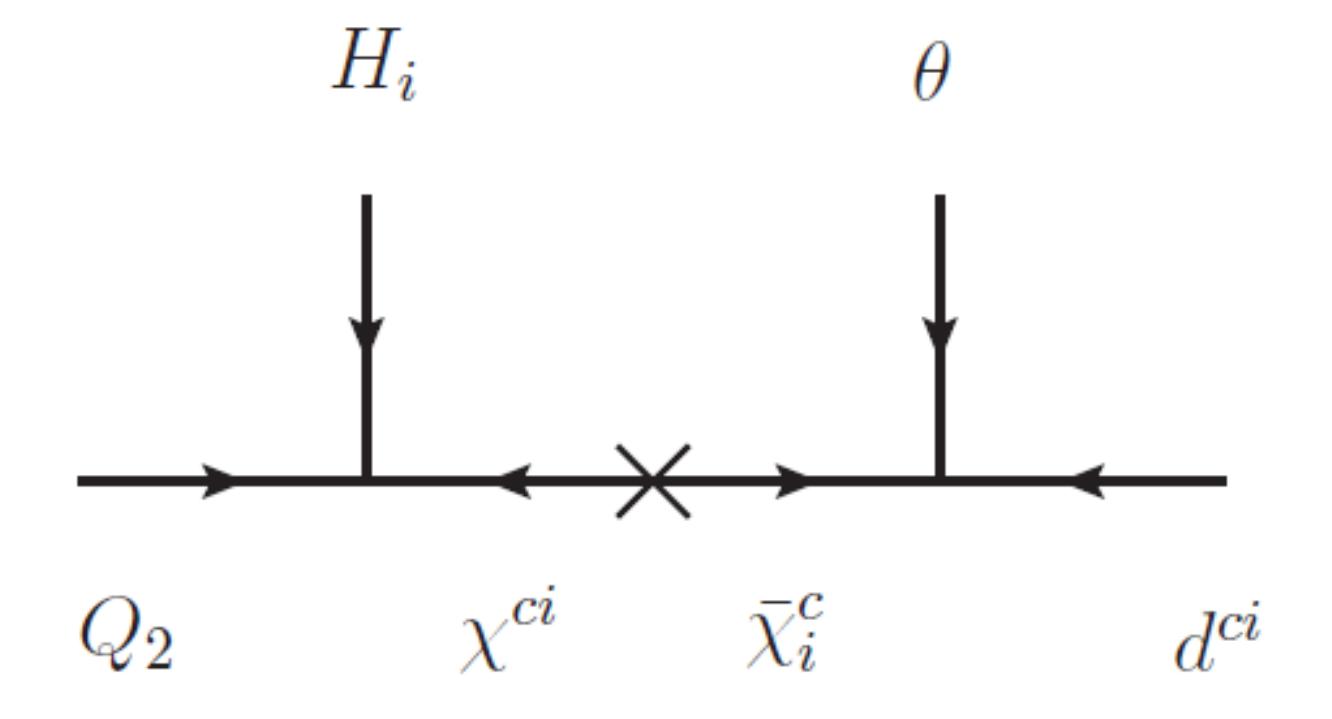}
\caption{Diagrams of possible $Q_2 (H d^c)_{01} \theta$ topologies. The $\chi$,$\bar{\chi}$ messengers are $SU(2)$ doublets whereas the $\chi^{ci}$, $\bar{\chi}^c_i$ are $\Delta(27)$ triplets.\label{fig:top}}
\end{figure}

The $3_{01} \times 3_{02}$ product involving $H$ (or $H^\dagger$) produces the corresponding singlet $1_{00}$, $1_{01}$ or $1_{02}$ matching each of the $Q_i$ combinations, and the invariants in the quark sector are of the generic form $Q_i (H^\dagger u^c)$ and $Q_i (H d^c)$.
It was already mentioned that one of the goals is to alleviate the hierarchical Yukawa couplings required by the observed fermion masses. As we are adding a flavour symmetry to the SM, leaving the hierarchies unexplained is unappealing. To tackle this issue, we use a generalised FN mechanism where for each generation there is a specific $U(1)_F$ or $Z_N$ charge that must be cancelled by some additional $\theta$, $\varphi$ insertions in order to build invariant terms.
With the content listed in Table \ref{ta:content} the specific invariants are:
\begin{align}
L_d = y_3 Q_3 (H d^c)_{01} + &y_2 Q_2 (H d^c)_{00} \varphi^2 + y_1 Q_1 (H d^c)_{00} \varphi^3 \\
+&p_{2} Q_2 (H d^c)_{01} \theta + p_{1} Q_1 (H d^c)_{01} \varphi \theta\\
+h_{3} (H H^\dagger)  Q_3 (H d^c)_{01} + &h_{2} (H H^\dagger)  Q_2 (H d^c)_{00} \varphi^2 + h_{1} (H H^\dagger)  Q_1 (H d^c)_{00} \varphi^3.
\label{eq:Ld}
\end{align}
The mass scale suppressions required for the ultraviolet completion of each non-renormalisable terms, have been absorbed into the definition of the respective $y_i$, $p_i$ and $h_i$ coefficients and are otherwise omitted for simplicity. In a complete model they can be identified as the masses of the suitable FN messenger fields. Two alternatives for the $Q_2 (H d^c)_{01} \theta$ term are shown in Figure \ref{fig:top} for the sake of illustration.

If one is using $U(1)_F$, the only other allowed non-renormalisable invariants require adding to each of the already invariant terms an independently $U(1)_F$ invariant combination such as $\theta^\dagger \theta$ or $\theta^\dagger \varphi^2$, e.g. $Q_3 (H d^c)_{02} \theta^\dagger \varphi^2$. The first type of combination is also $\Delta(27)$ invariant and merely redefines the existing terms, and the second type only appears when suppressed by three (or more) additional field insertions.
In turn, with a discrete symmetry, the lowest viable $N$ for this specific framework is $4$ (otherwise $Q_1$ becomes neutral) and the equivalent non-negative assignments are $3$ for $\varphi$ and $2$ for $\theta$, so there would be additional invariants with two additional insertions $\theta^2$ e.g. $Q_3 (H d^c)_{02} \theta \theta$.
This third type of invariants is only allowed by the discrete symmetry, and can be pushed to higher orders simply by using a larger $Z_N$ group.

As seen in eq.~(\ref{eq:Ld}), the field $\theta$ ($1_{02}$) combines with $Q_1$, $Q_2$ (both $1_{00}$) to respectively make the $p_1$, $p_2$ invariants of the down sector.
Because $Q_3$ transforms as $1_{02}$, there are $(H d^c)_{01}$ invariants involving all three $Q_i$.
Therefore, the respective $M_d$ structure has in each row one contribution with the $\omega$ coming from the $\langle H_1 \rangle$ VEV in the same column (the third column):
\begin{equation}
 M_d = M + M_p + M_h
\end{equation}
\begin{equation}
M = v \begin{pmatrix}
	y_{1} \omega & y_{1} & y_{1} \\
	y_{2} \omega & y_{2} & y_{2} \\
	y_{3} & y_{3} & y_{3} \omega 
\end{pmatrix} \,,
\end{equation}
\begin{equation}
M_p = v \begin{pmatrix}
	p_{1} & p_{1} & p_{1} \omega \\
	p_{2} & p_{2} & p_{2} \omega \\
	0 & 0 & 0 
\end{pmatrix} \,,
\end{equation}
\begin{equation}
M_h = v^3 \begin{pmatrix}
	h_{1} & h_{1} \omega^2 & h_{1} \omega^2 \\
	h_{2} & h_{2} \omega^2 & h_{2} \omega^2 \\
	h_{3} \omega^2 & h_{3} \omega^2 & h_{3} \,.
\end{pmatrix}
\end{equation}
This is sufficient to independently populate all off-diagonal entries of $M_d M_d^\dagger$ and generate the CKM matrix.
The $M_d M_d^\dagger$ $23$ and $31$ entries which in \cite{Bhattacharyya:2012pi} arose from $Q_3 (H d^c)_{00}$ interfering with $Q_{2} (H d^c)_{00}$, $Q_{2} (H d^c)_{00} (H H^\dagger)$ and $Q_{1} (H d^c)_{00}$, $Q_{1} (H d^c)_{00} (H H^\dagger)$ respectively, arise here instead from $Q_{2,1} (H d^c)_{01}$ interfering with $Q_{3} (H d^c)_{01}$ and $Q_{3} (H d^c)_{01} (H H^\dagger)$).\footnote{$M$ is rank 2, so 3 non-zero masses required contributions beyond $M$ as well. The same applies to the up quark sector.}

It is easy to check that without the structure associated with $M_h$ the complex phase present in $M_d$ does not survive into $M_d M_d^\dagger$. Thus the $(H H^\dagger)$ invariants are directly responsible for the CKM matrix being complex (as was the case already in \cite{Bhattacharyya:2012pi}). The structure shown in $M_h$ is the only new contribution coming from the possible $\Delta(27)$ invariants (the other contributions can be absorbed into the already existing structures, by making use of the $1 + \omega + \omega^2=0$ identity). 

In the up sector, due to the stronger hierarchy, the contributions to the CKM are negligible, and it is simple to fit the 3 masses with the leading Yukawa couplings:
\begin{align}
L_u=x_3 Q_3 (H^\dagger u^c)_{01} + &x_2 Q_2 (H^\dagger u^c)_{00} \varphi^2 + x_1 Q_1 (H^\dagger u^c)_{00} \varphi^3\\
+&q_{2} Q_2 (H^\dagger u^c)_{01} \theta + q_{1} Q_1 (H^\dagger u^c)_{01} \varphi \theta
\end{align}
Noting that $(u^c H^\dagger)_{01} = u^c_1 H^\dagger_3 + u^c_2 H^\dagger_1 + u^c_3 H^\dagger_2$ as the $3_{02}$ in the product is $H^\dagger$ (in contrast with
$(H d^c)_{01} = H_1 d^c_3 + H_2 d^c_1 + H_3 d^c_2$), we have:
\begin{equation}
M_u = v \begin{pmatrix}
	x_{1} \omega^2 & x_{1} & x_{1} \\
	x_{2} \omega^2 & x_{2} & x_{2} \\
	x_{3} & x_{3} \omega^2 & x_{3} 
\end{pmatrix}
+
v \begin{pmatrix}
	q_{1} & q_{1} \omega^2 & q_{1} \\
	q_{2} & q_{2} \omega^2 & q_{2}  \\
	0 & 0 & 0 
\end{pmatrix} \,.
\end{equation}

We now focus on the the full renormalisable scalar potential. The scalars are $H$, $\varphi$ and $\theta$:
\begin{align}
 V(H,\varphi, \theta) =&m_\varphi^2 \varphi \varphi^\dagger + m_\theta^2 \theta \theta^\dagger + m_H^2 \left[ H_i H_i^\dagger \right] \\
+&\lambda_\varphi (\varphi\varphi^\dagger)^2 +\lambda_\theta (\theta \theta^\dagger)^2 + \lambda_{\varphi \theta}(\varphi\varphi^\dagger)(\theta \theta^\dagger)+\lambda_1 \left[ (H_i H_i^\dagger)^2\right]\\
+&\lambda_2 \left( H_1 H_1^\dagger H_2 H_2^\dagger + H_2 H_2^\dagger H_3 H_3^\dagger + H_3 H_3^\dagger H_1 H_1^\dagger \right)\\
+&\lambda_3 \left( H_1 H_2^\dagger H_1 H_3^\dagger + H_2 H_3^\dagger H_2 H_1^\dagger +H_3 H_1^\dagger H_3 H_2^\dagger +\text{h.c.} \right)\\
+&\left(\lambda_{\varphi H} \varphi \varphi^\dagger +  \lambda_{\theta H} \theta \theta^\dagger \right) \left[ H_i H_i^\dagger \right] 
\,,
\end{align}
where $\lambda_3>0$ in eq.~(\ref{eq:VEV}) is responsible for the GCPV solution - see \cite{Branco:1983tn} (or \cite{Varzielas:2012pd,Varzielas:2013zbp} for details). The phase is not affected by terms involving $\theta \theta^\dagger$ as this combination transforms as a trivial $\Delta(27)$ singlet and can not enable any additional phase-dependent invariants. Any such terms that would compromise the GCPV solution would only appear at the non-renormalisable level - e.g. terms with $H H^\dagger$ and a single $\theta$ would require at least an additional $(\varphi^\dagger)^2$ in order to be invariant under the extra $U(1)_F$ symmetry.
At this point one can see that the charges of the $\Delta(27)$ singlets must be different, otherwise the renormalisable term $(H H^\dagger) (\varphi \theta^\dagger)$ would be allowed and phase-dependent, spoiling GCPV.

As the masses and quartic couplings of any potential terms involving either $\varphi$ or $\theta$ (or both) are entirely separated from the terms involving $H$, it is natural to have the scales involved in determining their respective VEVs be different and larger than the scale associated with breaking the electroweak symmetry.
There is therefore no tension from using $\varphi$ or $\theta$ as the scalars in the FN mechanism that generates hierarchies in the quark Yukawa couplings when suitable messengers are introduced (see e.g. Figure \ref{fig:top}). We do not delve here into further details of the messenger sector.

As described in \cite{Bhattacharyya:2012pi}, it is easy to accommodate with the parameters of the scalar potential a situation of decoupling, where the additional $\Delta(27)$ singlet scalars have masses of order TeV. Within this limit, one can obtain without fine-tuning the parameters that only one of the CP even states associated with $H$ is light, playing the role of the SM Higgs (at 125-126 GeV). There is however a more interesting scenario which is also viable, where two CP even states are light:  $h_a$, that avoids detection due to not having $h_a V V$ couplings with the SM vector bosons $V$ and $h_b$ at at 125-126 GeV and SM-like couplings, only $h_c$ being heavier.


In summary, although \cite{Bhattacharyya:2012pi} showed that geometrical CP violation in a $\Delta(27)$ framework could be made compatible with the precision flavour data, it had two problems: it ignored the hierarchy of the Yukawa couplings, and the required non-trivial $\Delta(27)$ singlet jeopardised the calculable phase.
Here, we go towards a more realistic model and address both issues by enlarging the symmetry and field content to include an additional symmetry (either an $U(1)_F$ or a discrete $Z_N$) and two scalars charged under it.
One of these two scalars is a non-trivial $\Delta(27)$ singlet that replaces the one used in \cite{Bhattacharyya:2012pi}, and no longer poses a threat to geometrical CP violation as it is charged under the extra symmetry.
Both scalars acquire a VEV that is naturally larger than the electroweak scale, and by suitably charging the lighter generations of the quark doublets under the same symmetry, a variant of the Froggatt-Nielsen mechanism is enacted to explain the Yukawa hierarchies.
The interesting properties of the scalar sector described in \cite{Bhattacharyya:2012pi} are preserved.

IdMV was supported by DFG grant PA 803/6-1 and by the Swiss National Science Foundation.
DP was supported by DFG grant PA 803/5-1.

\bibliography{v2}
\bibliographystyle{apsrev4-1}

\end{document}